\documentclass[aps, prd, twocolumn, superscriptaddress, nofootinbib]{revtex4}
\usepackage{graphicx}
\usepackage{dcolumn}
\usepackage{amssymb}
\usepackage{amsmath,amssymb,amsfonts}

\begin{document}

\newcommand{\Eq}[1]{\mbox{Eq. (\ref{eqn:#1})}}
\newcommand{\Fig}[1]{\mbox{Fig. \ref{fig:#1}}}
\newcommand{\Sec}[1]{\mbox{Sec. \ref{sec:#1}}}

\newcommand{\PHI}{\phi}
\newcommand{\PhiN}{\Phi^{\mathrm{N}}}
\newcommand{\vect}[1]{\mathbf{#1}}
\newcommand{\Del}{\nabla}
\newcommand{\unit}[1]{\;\mathrm{#1}}
\newcommand{\x}{\vect{x}}
\newcommand{\y}{\vect{y}}
\newcommand{\p}{\vect{p}}
\newcommand{\ScS}{\scriptstyle}
\newcommand{\ScScS}{\scriptscriptstyle}
\newcommand{\xplus}[1]{\vect{x}\!\ScScS{+}\!\ScS\vect{#1}}
\newcommand{\xminus}[1]{\vect{x}\!\ScScS{-}\!\ScS\vect{#1}}
\newcommand{\diff}{\mathrm{d}}
\newcommand{\mk}{{\mathbf k}}
\newcommand{\ep}{\epsilon}

\newcommand{\be}{\begin{equation}}
\newcommand{\ee}{\end{equation}}
\newcommand{\bea}{\begin{eqnarray}}
\newcommand{\eea}{\end{eqnarray}}
\newcommand{\vu}{{\mathbf u}}
\newcommand{\ve}{{\mathbf e}}
\newcommand{\vn}{{\mathbf n}}
\newcommand{\vk}{{\mathbf k}}
\newcommand{\vz}{{\mathbf z}}
\newcommand{\vx}{{\mathbf x}}
\def\dup{\;\raise1.0pt\hbox{$'$}\hskip-6pt\partial\;}
\def\ddn{\;\overline{\raise1.0pt\hbox{$'$}\hskip-6pt\partial}\;}


\title{Primordial standing waves}
\newcommand{\addressImperial}{Theoretical Physics Group, The Blackett Laboratory, Imperial College, Prince Consort Rd., London, SW7 2BZ, United Kingdom}
\newcommand{\addressRoma}{Dipartimento di Fisica, Universit\`a di Roma ``La Sapienza'', P.le A. Moro 2, 00185 Roma, Italy}
\newcommand{\addressRadboud}{Radboud University, Institute for Mathematics, Astrophysics and Particle Physics, Heyendaalseweg 135, NL-6525 AJ Nijmegen, The Netherlands}

\author{Giulia Gubitosi}
\affiliation{\addressRadboud}\affiliation{\addressRoma}
\author{Jo\~{a}o Magueijo}
\affiliation{\addressImperial}

\date{\today}

\begin{abstract}
We consider the possibility that the primordial fluctuations (scalar and tensor) might have been standing waves at their moment of creation, whether or not they had a quantum origin. We lay down the general conditions for spatial translational invariance, and isolate the pieces of the most general such theory that comply with, or break translational symmetry. We find that, in order to characterize statistically translationally invariant standing waves, it is essential to consider the correlator 
$\langle c_0(\vk) c_0(\vk ')\rangle$ in addition to the better known
$\langle c_0(\vk) c_0^\dagger(\vk ')\rangle$ (where $c_0(\vk)$ are the complex amplitudes of travelling waves).  We then examine how the standard process of ``squeezing'' (responsible for converting travelling waves into standing waves while the fluctuations are outside the horizon) reacts to being fed primordial standing waves. For translationally invariant systems only one type of standing wave, with the correct temporal phase (the ``sine wave''), survives squeezing. Primordial standing waves might therefore be invisible at late times -- or not -- depending on their phase. Theories with modified dispersion relations behave differently in this respect, since only standing waves with the opposite temporal phase survive at late times.
\end{abstract}

\keywords{cosmology}
\pacs{}

\maketitle

\section{Introduction}

It is well known that at late times primordial density fluctuations must form standing waves, as opposed to traveling waves \cite{grish, pedro}. Regardless of how they were initially produced and found themselves outside the horizon, they must reenter the horizon as standing waves with a specific temporal phase in order to comply with observational constraints \cite{Dodelson:2003ip}. This may happen for a variety of reasons, with ``squeezing'' usually taking the blame. The matter was reexamined in~\cite{us}, both from the phenomenological point of view and in the context of several scenarios, inflationary and otherwise. 

But what if the {\it primordial} mechanism responsible for the fluctuations were to generate standing, instead of travelling waves? This possibility was first considered in~\cite{PRL} in relation to the coupling between right and left gravitons due to the  non-trivial inner product in Hilbert space present in some quantum gravity theories. Putting aside the issue of chirality, it is evident that if the coupled waves move in opposite directions, then their correlations would produce standing waves (a matter recognized in~\cite{PRL}). However, the correlators considered in~\cite{PRL} break translational invariance, something which is by no means necessary for a general process producing standing waves, as we shall see. In addition, in~\cite{PRL} the primordial standing waves were not followed up until horizon reentry, to investigate how ``squeezing'' reacts to the input of waves which are already standing waves. 

In this short paper we rectify these two matters. Given that primordial fluctuations are usually studied in terms of the correlation matrix of travelling waves, in Section~\ref{gym} we start by examining how standing waves may be described in terms of constrained travelling waves moving in opposite directions. 
Then, in Section~\ref{transl} we study how statistical translational invariance can be implemented for a stochastic ensemble of travelling and standing waves. It turns out that for a given headless direction, we need to give uniform distributions to two phases in the case of travelling waves, whereas only one phase needs to be suitably randomized in the case of standing waves. In Section~\ref{corrk-k} we consider the most general correlation matrix for travelling waves moving in opposite directions, and examine the pieces which do and do not break translational invariance, and those that are related to standing wave production. 
By assuming that amplitudes and phases are uncorrelated we are able to evaluate this matrix in general.

Finally, in Section~\ref{squeeze} we study how primordial standing waves interact with the usual processes (``squeezing" or otherwise) responsible for converting primordial traveling waves into late-time standing waves. We find that the temporal phase of the primordial standing wave is crucial for the transmission rate. Waves with the same temporal phase as those that would be produced by squeezing in the standard scenario are enhanced (the ``sine wave''), whereas waves with complementary phase (the ``cosine wave'') are suppressed. The opposite result is obtained  if instead of squeezing (e.g. in the inflationary or bimetric VSL scenario~\cite{Jprd}) one considers the evolution of fluctuations in theories with modified dispersion relations~\cite{Amelino-Camelia:2013tla}.

\section{The gymnastics of travelling and standing waves} \label{gym}
In~\cite{us} we reviewed in detail the differences between standing and travelling waves, and their inter-relation. A standing wave has a (real) expression where space and time oscillations can be factored:
\be\label{standing}
y_{st.}=A(\overline \vk)\cos(\omega\eta +\phi_{\vk t})\cos(\vk\cdot \vx +\phi_{\vk x}).
\ee
As such its spatial nodes are fixed, so that the wave is ``standing'' ($\overline\vk$ denotes a headless vector, since the ``sign'' of $ \vk$  is therefore irrelevant). Instead, for a travelling wave this factorization cannot be achieved in the real domain (although it can be done using complex functions and then taking the real part), with the wave usually written as:
\be\label{travelling}
y_{tr.}=B( \vk)\cos(-\omega\eta+\vk\cdot \vx +\phi_\vk).
\ee
For a travelling wave the spatial nodes move in time, and  the ``sign'' of $\vk$ is relevant, since it denotes the direction of motion.  

Assuming ${\vk}$ is fixed (and thus $\omega$, via the dispersion relation), a travelling wave has 2 real degrees of freedom (d.o.f): the real amplitude $B(\vk)$ and the phase $\phi_\vk$ (often packaged as the complex amplitude $Be^{i\phi_\vk}$ when using complex notation; see below). Instead, fixing ${\overline \vk}$, a standing wave has 3 d.o.f.: the real amplitude $A(\overline \vk)$ and 2 phases, one spatial, $\phi_{\vk x}$, and one temporal, $\phi_{\vk t}$. Thus, for fixed ${\overline \vk}$ the space of all travelling waves  is 4 dimensional (2 d.o.f for each of $\vk$ and $-\vk$), whereas the space of all standing waves is 3 dimensional.

One can always build a standing wave from two traveling waves with the same real amplitude and $\overline\vk$, moving in opposite directions. This matches the counting of d.o.f, since $4-1=3$, where the $-1$ discounts the constraint enforcing equal real amplitudes. Specifically, setting $B(\vk)=B(-\vk)$ and denoting by $\phi_{\pm\vk}$ the phases of the two travelling waves, one finds~\cite{us} that:
\bea\label{travelling1}
y &=& B( \vk)\cos(-\omega\eta+\vk\cdot \vx +\phi_\vk)\nonumber\\
&+&B(- \vk)\cos(\omega\eta+\vk\cdot \vx - \phi_{-\vk}),
\eea
equals a standing wave of form (\ref{standing}), with  
 $A(\overline \vk)=2B(\vk)$ and 
\bea
\phi_{\vk t}&=&-\frac{\phi_\vk + \phi_{-\vk}}{2}\label{phit}\\
\phi_{\vk x}&=&\frac{\phi_\vk -\phi_{-\vk}}{2}.\label{phix}
\eea
(reference \cite{us} used the alternative notation $\phi_\pm=\pm \phi_{\pm\vk}$; we shall presently justify our notation). These expressions relating the phases of standing and travelling waves will be essential in discussing statistical translational invariance for standing waves.

In general the space of all possible oscillatory solutions with fixed $\overline \vk$ is 4 dimensional.
In fact, it can be spanned by the most general superposition of travelling waves moving along $\pm \vk$ (counting d.o.f.,  $4=2+2$). This is usually written splitting the momentum space in one half and writing modes moving in opposite directions as part of the complex amplitude of each Fourier mode for half the space. More concretely, using complex notation we have:
\be\label{FT1}
y(\vx)=\int \frac{d\vk}{(2\pi)^3} \, e^{i\vk\cdot \vx}y(\vk)=
\int_{{\rm I\!R}_3^+} \frac{d\vk}{(2\pi)^3} \, e^{i\vk\cdot \vx}y(\vk)+ c.c.
\ee
with 
\be\label{solrad0}
y(\vk)=\frac{e^{-i\omega \eta}}{\sqrt{2\omega}} c_0(\vk)+
\frac{e^{i\omega \eta}}{\sqrt{2\omega }}c_0^\dagger (-\vk),
\ee
where $c_0(\pm\vk)$ denotes the complex amplitude of the mode moving along $\pm\vk$.
With some algebra we then get the dictionary between real (Eq.~\ref{travelling}) and complex notation:
\be\label{compamp}
c_0(\vk)=\sqrt{\frac{\omega}{2}}B(\vk)e^{i\phi_\vk}.
\ee
Thus, we have packaged into a complex amplitude the real amplitude and the phase. We have also included the two traveling modes in the single amplitude of the Fourier mode $y(\vk)$, with the convention that the sign of $\pm\vk$ is relevant and points to the direction of propagation of the wave. With this notation we will be able to bridge better the standard theory of cosmological perturbations and the generation of standing waves.

\section{Translational invariance for independent standing and travelling waves}\label{transl}
We now examine how statistical translational invariance can be enforced, or broken, for an  ensemble of 
waves, separating the case of standing and travelling waves. We consider first a single mode, i.e we fix $\overline \vk$ for a standing wave, and $\vk$ for a travelling wave. How can we randomize the 3 d.o.f. of a standing wave, or the 2 d.o.f. of such a travelling wave, in a way that complies with translational invariance?

As can be read off from (\ref{travelling}), a translation of the coordinate system:
\be\label{xtrans}
\vx\rightarrow \vx + \Delta\vx
\ee
shifts the phase of a travelling wave by\footnote{The convention adopted for $\phi_{\pm\vk}$ (which differs
from the $\phi_\pm$ used in~\cite{us}), ensures that this formula is valid for both $\pm\vk$.}:
\be\label{deltaphi}
\phi_\vk\rightarrow \phi_\vk + \vk\cdot \Delta\vx.
\ee
Thus, the only way in which a distribution of travelling waves can be translationally invariant is for it not to depend on the phase of the wave. Statistically translational invariant random travelling waves must, therefore, have a uniform distribution for the marginal distribution of their phases. 

Standing waves, in contrast, have two types of phases (spatial and temporal) and, as can be read off from (\ref{standing}),  under spatial translation (\ref{xtrans}) these transform as:
\bea
\phi_{\vk x}&\rightarrow &\phi_{\vk x} + \vk\cdot \Delta\vx\\
\phi_{\vk t}&\rightarrow &\phi_{\vk t}\, .
\eea
Therefore, only the spatial phase needs to be uniformly distributed in a translationally invariant ensemble. If they were to be required to be invariant under time translations, one would need to give a uniform distribution to $\phi_t$ too, but this is often not assumed in cosmology.

These considerations apply to each of the modes separately, whether they are independent or not (if they are not independent, the statements made refer to the marginal distributions of each mode with respect to all others). They also apply directly to the modes' joint distributions, if they are statistically independent. This is usually a common assumption for travelling waves. 

However, as we saw in Section~\ref{gym}, a standing  wave may be seen as two travelling waves with the same amplitude moving in opposite directions. Primordial production of fluctuations is usually discussed in terms of travelling waves. Therefore, if we wish to discuss primordial production of standing waves in terms of travelling waves we have to be able to introduce at least correlations between $\vk$ and $-\vk$. This is what we shall do in the next Section, where we gain a better understanding of how translational invariance for travelling and standing waves interact.

\section{Translational invariance for correlated waves moving in opposite
 directions} \label{corrk-k}
Let us take half of Fourier space, ${{\rm I\!R}_3^+}$, pick a generic $\vk$, and consider 
the most general correlation matrix for travelling waves moving along $\vk$ and $-\vk$. It is standard
in the study of production of primordial fluctuations to consider this matrix in terms of 
complex amplitudes (\ref{compamp}), imposing the diagonal form:
\be\label{cordiag}
\langle c_0(\vk) c_0^\dagger(\vk ')\rangle=\delta(\vk-\vk')P(k).
\ee
Correlations between different $\vk$ (and in particular between $\vk$ and $-\vk$) are usually ruled out  on the grounds of translational invariance. The last statement is actually incorrect, indeed were it correct  it would preclude the production of translationally invariant standing waves (built from correlations between $\vk$ and $-\vk$). In order to capture translationally invariant correlations, however, we must add to (\ref{cordiag}) another independent correlator: $\langle c_0(\vk) c_0(\vk ')\rangle$.

We shall therefore consider the more general matrix:
\bea\label{corrmat}
\langle c_0(\vk) c_0^\dagger(\vk ')\rangle&=&\delta(\vk-\vk')P(\vk)+\delta(\vk+\vk')W(\vk)\;\;\;\label{cormat1}\\
\langle c_0(\vk) c_0(\vk ')\rangle&=&\delta(\vk-\vk')Z(\vk)+\delta(\vk+\vk')Q(\vk).\;\;\;\label{cormat2}
\eea
Standing waves production is signalled by correlations between $\vk$ and $-\vk$ modes, and so by non-vanishing $Q$ or $W$. However these two components are very different in nature. 
Under translation (\ref{xtrans}) we have (see \eqref{compamp}):
\be
c_0(\vk)\rightarrow c_0(\vk)e^{i\vk\cdot\Delta\vx},
\ee
so that we find transformation laws:
\bea
P(\vk)&\rightarrow&P(\vk)\\
W(\vk)&\rightarrow&W(\vk)  e^{2i\vk\cdot\Delta\vx} \\
Z(\vk)&\rightarrow& Z(\vk) e^{2i\vk\cdot\Delta\vx}  \\
Q(\vk)&\rightarrow&Q(\vk).\label{Qlabel}
\eea
Therefore $W$ is associated with the production of standing waves with preferential positions for the nodes, breaking translational invariance. Such waves were already considered in~\cite{PRL}, and this process gives $\phi_{\vk x}$ a non-uniform distribution, as explained in Section~\ref{transl}. 
However, translational symmetry  breaking is not necessary when producing standing waves. 
As (\ref{Qlabel}) shows, $Q\neq 0$ (and $W=0$) is associated with the production of translationally invariant standing waves. Such waves have random uniformly distributed $\phi_{\vk x}$, but not $\phi_{\vk t}$. In order to capture this process it is essential to consider the correlator (\ref{cormat2}), which is independent from (\ref{cormat1}). This has been omitted in previous literature. 
In contrast, we abstain from writing the correlator $\langle c_0^\dagger (\vk) c_0^\dagger(\vk ')\rangle$ because this is simply the complex conjugate of $\langle c_0(\vk) c_0(\vk ')\rangle$. 
(Parenthetically we note that $Z$ measures the breaking of translational invariance for purely uncorrelated travelling waves.) 

We remark that 
in writing the correlation matrix (\ref{cormat1}) and (\ref{cormat2}) we can assume isotropy where appropriate, and so we can drop the direction of $\vk$ from the argument of $P(\vk)$ and $Q(\vk)$. For the components which break translational invariance it may be questionable to assume isotropy (it would require that we be at the centre of the universe). However this is possible, with the proviso that, for example,  $W$ would then have to be real (since  $W(-\vk)=W^\star(\vk)$), limiting the possible preferred nodal positions. A similar remark  applies to $Z$.
In general all the components of the correlator, with the exception of $P(k)$, are complex, with the phases containing important information.

In order to understand how the various components of the correlators relate to the phases of the waves
let us write the correlators in terms of real amplitudes and phases, as in (\ref{compamp}), assuming that the two are independent random variables. We find:
\bea
P(k)&=&\frac{\omega}{2}{\langle B^2(k) \rangle}\\
 W(\vk)&=&\frac{\omega}{2}{\langle B(\vk) B(-\vk)\rangle}\langle e^{i(\phi_\vk-\phi_{-\vk})}\rangle \\
 Z(\vk)&=&\frac{\omega}{2}{\langle B^2(k) \rangle}\langle e^{2i\phi_\vk}\rangle \\
 Q(k)&=&\frac{\omega}{2}{\langle B(\vk) B(-\vk) \rangle}\langle e^{i(\phi_\vk+\phi_{-\vk})}\rangle \, .
\eea
For independent traveling waves $W=Q=0$, and  in order to impose translational invariance we should uniformly randomize the two phases, $\phi_\vk$ and $\phi_{-\vk}$, so that $Z=0$. For pure standing waves production ${\langle B(\vk) B(-\vk)\rangle}={\langle B^2(k) \rangle}$ and what we do about the phases is now completely different. 
If there are perfect correlations, we only have to uniformly randomize the difference $\phi_\vk-\phi_{-\vk}$. This renders $W=0$. The sum of the phases, instead, does not need to be uniformly distributed to comply with translational invariance, and indeed it can be fixed (as is the case with waves after squeezing). Under a translation we have in fact
\be
\phi_\vk+\phi_{-\vk}\rightarrow 
\phi_\vk+\phi_{-\vk}.
\ee
Giving this quantity a non-uniform distribution is equivalent to doing the same to the temporal phase $\phi_{{\vk }t}$ of standing waves. 
Thus, for independent travelling waves, we have to give uniform distributions to two phases separately, whereas for standing waves only one phase needs to be randomized.

\section{Transfer of standing waves through standard ``squeezing''}\label{squeeze}

In the usual inflationary scenario (and in bimetric VSL theories \cite{Jprl, Jprd}) squeezing is the mechanism responsible for turning independent travelling waves into standing waves with a sine temporal phase \cite{us}. Let us now consider what happens if standing waves, rather than independent travelling waves, are fed into the usual mechanism of squeezing.  Then, after the modes leave the horizon the solution takes the approximate form \cite{us}:
\be\label{late-sol}
y\approx \frac{-i}{\sqrt {2}k^{3/2}\eta}(c_0(\vk)-c_0^\dagger(-\vk)).
\ee
Assuming translational invariance (i.e. setting $Z=W=0$) we can therefore 
use (\ref{cormat1}) and (\ref{cormat2}) to compute:
\be
{\langle y(\vk)y^\star (\vk')\rangle}=\frac{\delta(\vk-\vk')}{k^3\eta^{2}}
[P(k)-\Re Q(k)]
\ee
stressing the physical importance of the complex phase of $Q$. If the process producing the primordial fluctuations led to perfect standing waves we would have:
\be
Q(k)=P(k){\langle e^{-2i\phi_{kt}}\rangle}
\ee
(see the computation at the end of the last Section). This is a general formula, but
in many scenarios 
the distribution of $\phi_{kt}$ is a delta function, with a peak at a value $\phi_t$ independent of $k$. In that case, 
\be
{\langle y(\vk)y^\star(\vk')\rangle}=\frac{\delta(\vk-\vk')}{k^3\eta^{2}}
P(k)(1-\cos (2\phi_t)).
\ee
We see that standing waves with $\phi_t=0$ disappear after squeezing, becoming invisible at late times. 
This should be obvious from (\ref{late-sol}), since for such waves $\phi_\vk=-\phi_{-\vk}$, and so $c_0(\vk)=c^\dagger (-\vk)$.  These waves are cosine waves in time, and so complementary to those produced by squeezing in the standard scenario.

Waves with the $\phi_t=\pi/2$, instead are amplified by a factor of 2. These are the standing waves with the same time phase as those produced at the end of squeezing, i.e. sine waves in time. For the intermediate angle $\phi_t=\pi/4$ the end product is the same as if independent travelling waves had been inputed.

This is by no means a general conclusion in  phenomenologically viable theories. 
As a counter-example of the findings just discussed, let us consider theories with modified dispersion relations (MDR), of the kind that is known to produce scale-invariant perturbations \cite{Amelino-Camelia:2013tla, Amelino-Camelia:2013wha, Amelino-Camelia:2013gna, Amelino-Camelia:2015dqa, Horava}.
In this case, the solution for primordial perturbations after horizon exit is \cite{us, usMDR}:
\be
y\approx \frac{1}{\sqrt{2\omega}}(c_0(\vk)+c_0^\dagger(-\vk))\,.
\ee
Following  steps similar to those described above, we find the correlator:
\bea
{\langle y(\vk)y^\star (\vk')\rangle}&=&\frac{\delta(\vk-\vk')}{ \omega}
[P(k)+\Re Q(k)] \nonumber\\
&=& \frac{\delta(\vk-\vk')}{ \omega}
P(k)(1+\cos (2\phi_t))  \,,
\eea
where in the second step we assumed that pure standing waves are produced in the primordial universe, and that their phases are independent of $k$.
Thus, we see that within MDR models, primordial standing waves with cosine temporal  phase  are amplified, while waves with a sine temporal phase are obliterated at late times. This is one of the few aspects in which MDR scenarios can be distinguished from inflation and other scenarios (but see also~\cite{usMDR}).

\section{Conclusions}
Observational constraints on the Doppler peaks in the cosmic microwave background power spectrum show that primordial perturbations enter the horizon at late times as standing waves with a sine temporal phase.
It is usually assumed that perturbations are formed as stochastic traveling waves and then some mechanism (such as squeezing in the inflationary scenario) turns them into standing waves.
In this paper we examined the possibility that the primordial fluctuations were already standing waves before they left the horizon. This possibility could arise from correlations between modes moving in opposite directions, due to a non-trivial inner product structure in the quantum gravitational Hilbert space. Indeed such correlations were first suggested in~\cite{PRL} in relation to the possible coupling between right and left gravitons.
In that context the correlations  and the resulting standing waves  break translational invariance. 

In this paper we laid down  the conditions that an ensemble of standing waves has to satisfy in order not to break translational invariance: the spatial phases need to be uniformly distributed.  This observation allowed us to  show that a nontrivial correlator between the complex amplitudes of the primordial modes  can in fact encode translationally-invariant standing waves. This requires consideration of  the correlator  (\ref{cormat2}) between $c_{0}(\vk)$ and $c_{0}(\vk')$, besides the standard correlator between $c_{0}(\vk)$ and $c_{0}^{\dagger}(\vk')$.
We showed that, in the correlator between $c_{0}(\vk)$ and $c_{0}^{\dagger}(\vk')$, a non-zero contribution proportional to $\delta(\vk+\vk')$ (whose coefficient we called $W$) signals the production of standing waves with preferred nodal positions (and non-uniformly distributed $\phi_{\vk x}$). The same kind of contribution showing up in the correlator between $c_{0}(\vk)$ and $c_{0}(\vk')$ (whose coefficient we called $Q$) signals the production of standing waves with statistically translational invariant random nodal positions (and uniformly distributed $\phi_{\vk x}$). Finally, the contribution to the latter correlator coming from $\delta(\vk-\vk')$ (whose coefficient we called $Z$) signals breaking of translational invariance for travelling waves.

We then examined how the input of standing waves into the usual squeezing process would work out. Depending on the temporal phase of the standing waves, squeezing can either enhance or suppress them. Specifically, waves with a cosine temporal phase are suppressed, while those with a sine temporal phase are enhanced. The complementary result is found when instead of squeezing in the standard inflationary scenario one looks at the late-time behaviour of perturbations in theories with modified dispersion relations.

It is interesting that the non-trivial correlators responsible for primordial standing waves are in fact the 
``pump terms'' appearing in the squeezing formalism~\cite {Grishchuk:1992tw}. This is by now means accidental, since formally the two processes are very similar, even though in the former standing waves result from the structure of the inner product in Hilbert space, whereas in the latter from the structure of the Hamiltonian as the modes leave the horizon. But it does not mean the two could not interact non-trivially beyond what has been investigated in this paper.  In this sense it would be interesting to examine the impact of squeezing on the Ashtekar formalism~\cite{ash}, where the reality conditions intervene at a rather late stage in the quantization procedure to select the physical states as the ones we know~\cite{joaoash1,joaoash2}.

\section{Acknowledgements}

We thank Robert Brandenberger, Carlo Contaldi and Toby Wiseman for discussions related to this paper.  We acknowledge partial support from the John Templeton Foundation. JM was also supported by  an STFC consolidated  grant.


\end{document}